\documentclass[aps,preprint,amsmath,amssymb,superscriptaddress,nofootinbib]{revtex4}
\usepackage{multirow}
\usepackage{graphicx}
\usepackage{float}
\usepackage{mathtools}
\usepackage{array}
\newcolumntype{P}[1]{>{\centering\arraybackslash}p{#1}}
\begin{document}

\title{Search for the anomalous $WW\gamma$ couplings through the process $e^-e^+\,\rightarrow\,\nu_e\overline{\nu}_e\gamma$ at ILC with unpolarized and polarize beams}

\author{S. Spor}
\email[]{serdar.spor@beun.edu.tr}
\affiliation{Department of Medical Imaging Techniques, Zonguldak B\"{u}lent Ecevit University, 67100, Zonguldak, Turkey.}

\author{M. K{\"o}ksal}
\email[]{mkoksal@cumhuriyet.edu.tr} 
\affiliation{Department of Optical Engineering, Sivas Cumhuriyet University, 58140, Sivas, Turkey.} 

\begin{abstract}
We investigate the anomalous $W^+W^-\gamma$ couplings through the process $e^-e^+\,\rightarrow\,\nu_e\overline{\nu}_e\gamma$ for unpolarized and polarized electron (positron) beams at the International Linear Collider. We give the 95$\%$ Confidence Level limits on the anomalous couplings with and without the systematic uncertainties for various values of center-of-mass energies and the integrated luminosities. We show that the obtained limits on the anomalous couplings through the process $e^-e^+\,\rightarrow\,\nu_e\overline{\nu}_e\gamma$ can highly improve the current experimental limits. 
\end{abstract}


\maketitle

\section{Introduction}

The Standard Model (SM) of particle physics has proven a remarkably successful field theory at the electroweak scale and below. The gauge boson self-interactions are determined by the non-Abelian $SU(2)_L\times U(1)_Y$ gauge symmetry of the electroweak sector of the SM and also described by the triple gauge couplings (TGCs) such as $W^+W^-V$, $Z\gamma V$ and $ZZV$ $\left(V=\gamma, Z\right)$ \cite{Baur:2001ert,Yang:2013huy}. $W^+W^-V$ vertex involves charged couplings whereas $Z\gamma V$ and $ZZV$ involve neutral TGCs. Neutral TGCs at the tree level is forbidden due to lack of the electric charge of the Z boson. Neutral gauge boson self-couplings are permitted with loop diagrams in the SM. Therefore, studying the TGCs are of crucial importance to test the validity of the SM. Any deviation from the SM predictions would be a sign of the presence of new physics beyond the SM.  

The effective Lagrangian method is based upon the assumption that at higher energy regions beyond the SM, there is a more fundamental physics which reduces to the SM at lower energy regions. The model-independent approach via this effective Lagrangian method is used to investigate the new physics effect on $W^+W^-\gamma$ interactions. In this approach, in order to achieve effective interactions with SM particles, all heavy degrees of freedom are incorporated. 

We examine the effects of anomalous $W^+W^-\gamma$ couplings described with the effective Lagrangian method between W and $\gamma$ for the process $e^-e^+\,\rightarrow\,\nu_e\overline{\nu}_e\gamma$ at the International Linear Collider (ILC). New physics beyond the SM occurs with new interactions among the known particles. These new interactions contribute to the effective Lagrangian as higher dimensional operators, which are invariant under the SM symmetries and suppressed by the new physics scale $\Lambda$ \cite{Baak:2013yre}:

\begin{eqnarray}
\label{eq.1} 
{\cal L}_{eft}={\cal L}_{SM}+\sum_{d>4}\sum_i\frac{C_i}{\Lambda^{d-4}}{\cal O}_i\,,
\end{eqnarray}

{\raggedright where $d$ is the dimension of the operators. This effective Lagrangian reduces to the SM one in the limit $\Lambda\rightarrow\infty$. Since the coefficients of the higher dimensional operators, $C_i$, are fixed by the complete high energy theory, any extension of the SM can be parameterized by this effective Lagrangian, where $C_i$ are free parameters. Now, we will identify the effective Lagrangian of new physics including dimension-six operators that modify the interactions between electroweak gauge bosons:}

\begin{eqnarray}
\label{eq.2} 
{\cal L}_{eft}=\frac{1}{\Lambda^2}\left[C_W{\cal O}_W+C_B{\cal O}_B+C_{WWW}{\cal O}_{WWW}+C_{\tilde{W}WW}{\cal O}_{\tilde{W}WW}+C_{\tilde{W}}{\cal O}_{\tilde{W}}+h.c.\right]\,.
\end{eqnarray}

Only operators with even dimension can be constructed when baryon and lepton numbers are conserved. As a result, the largest contribution for new physics beyond the SM comes from dimension-six operators. Three CP-conserving dimension-six operators:

\begin{eqnarray}
\label{eq.3} 
{\cal O}_{WWW}=\text{Tr}\left[W_{\mu\nu}W^{\nu\rho}W_\rho^\mu\right]\,,
\end{eqnarray}
\begin{eqnarray}
\label{eq.4} 
{\cal O}_{W}=\left(D_\mu\Phi\right)^\dagger W^{\mu\nu}\left(D_\nu\Phi\right)\,,
\end{eqnarray}
\begin{eqnarray}
\label{eq.5} 
{\cal O}_{B}=\left(D_\mu\Phi\right)^\dagger B^{\mu\nu}\left(D_\nu\Phi\right)\,,
\end{eqnarray}

and two CP-violating dimension-six operators:

\begin{eqnarray}
\label{eq.6} 
{\cal O}_{\tilde{W}WW}=\text{Tr}\left[ \tilde{W}_{\mu\nu}W^{\nu\rho}W_\rho^\mu\right]\,,
\end{eqnarray}
\begin{eqnarray}
\label{eq.7} 
{\cal O}_{\tilde{W}}=\left(D_\mu\Phi\right)^\dagger \tilde{W}^{\mu\nu}\left(D_\nu\Phi\right)\,,
\end{eqnarray}

{\raggedright where $\Phi$ is the Higgs doublet field. The $D_\mu$ covariant derivative, $W_{\mu\nu}$ and $B_{\mu\nu}$ field strength tensors of the $SU(2)_I$ and $U(1)_Y$ gauge fields are respectively as follow:}

\begin{eqnarray}
\label{eq.8} 
D_\mu \equiv \partial_\mu\,+\,i\frac{g^\prime}{2}B_\mu\,+\,igW_\mu^i\frac{\tau^i}{2}\,,
\end{eqnarray}
\begin{eqnarray}
\label{eq.9} 
W_{\mu\nu}=\frac{i}{2}g\tau^i\left(\partial_\mu W_\nu^i - \partial_\nu W_\mu^i + g\epsilon_{ijk}W_\mu^j W_\nu^k\right)\,,
\end{eqnarray}
\begin{eqnarray}
\label{eq.10} 
B_{\mu\nu}=\frac{i}{2}g^\prime\left(\partial_\mu B_\nu - \partial_\nu B_\mu\right)\,,
\end{eqnarray}

{\raggedright where $\tau^i$ are the $SU(2)_I$ generators with Tr$[\tau^i\tau^j]=2\delta^{ij}$ $\left(i,j=1,2,3\right)$. $g$ and $g^\prime$ are $SU(2)_I$ and $U(1)_Y$ couplings, respectively. The effective Lagrangian for $W^+W^-\gamma$ interaction can be then parameterized by \cite{Li:2018tyb}:}

\begin{eqnarray}
\label{eq.11} 
{\cal L}_{WW\gamma}&=&ig_{WW\gamma}\Big[{g_1^{\gamma}}\left({W_{\mu\nu}^{+}}{W_{\mu}^{-}}A_{\nu}-{W_{\mu\nu}^{-}}{W_{\mu}^{+}}A_{\nu}\right) \nonumber \\
&+&\kappa_{\gamma}{W_{\mu}^{+}}{W_{\nu}^{-}}A_{\mu\nu}+\frac{\lambda_{\gamma}}{M_W^2}{W_{\mu\nu}^{+}}{W_{\nu\rho}^{-}}A_{\rho\mu} \nonumber \\
&+&ig_4^\gamma {W_{\mu}^{+}}{W_{\nu}^{-}}\left( \partial_\mu A_\nu+\partial_\nu A_\mu\right) \\
&-&ig_5^\gamma \epsilon_{\mu\nu\rho\sigma}\left({W_{\mu}^{+}}\partial_\rho{W_{\nu}^{-}}-\partial_\rho{W_{\mu}^{+}}{W_{\nu}^{-}}\right)A_\sigma \nonumber \\
&+&\tilde{\kappa}_{\gamma}{W_{\mu}^{+}}{W_{\nu}^{-}}\tilde{A}_{\mu\nu}+\frac{\tilde{\lambda}_{\gamma}}{M_W^2}{W_{\lambda\mu}^{+}}{W_{\mu\nu}^{-}}\tilde{A}_{\nu\lambda} \Big]\,, \nonumber 
\end{eqnarray}

{\raggedright where $g_{WW\gamma}=-e$ and $\tilde{A}=\frac{1}{2}\epsilon_{\mu\nu\rho\sigma}A_{\rho\sigma}$. $A^{\mu\nu}=\partial^\mu A^\nu - \partial^\nu A^\mu$ is the field strength tensor for photon.  In Eq.~(\ref{eq.11}), $g_1^{\gamma}$, $\kappa_{\gamma}$ and $\lambda_{\gamma}$ anomalous parameters are both C and P conserving while $g_4^\gamma$, $g_5^\gamma$, $\tilde{\kappa}_{\gamma}$ and $\tilde{\lambda}_{\gamma}$ anomalous parameters are C and/or P violating. Electromagnetic gauge invariance requires that ${g_1^{\gamma}}=1$. In the SM, the anomalous coupling parameters are given by $\kappa_{\gamma}=1$ ($\Delta \kappa_{\gamma}=0$) and $\lambda_{\gamma}=0$ at the tree level. However, CP-violating interactions can be confined individually to specially designed CP-odd observables that are insensitive to CP-even effects. Thus, the CP-conserving and violating interactions can be separated from each other. Here, the anomalous $\kappa_{\gamma}$ and $\lambda_{\gamma}$ coupling parameters can be reframed in terms of the couplings of the operators in Eq.~(\ref{eq.2}) and transformed into $c_{WWW}/\Lambda^2$, $c_{W}/\Lambda^2$ and $c_{B}/\Lambda^2$ \cite{Degrande:2013rry}. Thus, the effective field theory approach allows the following the relations between parameters:}

\begin{eqnarray}
\label{eq.12} 
{\kappa_\gamma}=1+\left(c_W+c_B\right)\frac{m_W^2}{2\Lambda^2}\,,
\end{eqnarray}
\begin{eqnarray}
\label{eq.13} 
{\lambda_\gamma}=c_{WWW}\frac{3g^2m_W^2}{2\Lambda^2}\,.
\end{eqnarray}

Similarly, the values of above $c_{WWW}/\Lambda^2$, $c_{W}/\Lambda^2$ and $c_{B}/\Lambda^2$ parameters lead to deviations from the SM for $W^+W^-\gamma$ couplings and determine new physics contributions. In the SM, the anomalous coupling parameters are given by $c_{WWW}/\Lambda^2=c_{W}/\Lambda^2=c_{B}/\Lambda^2=0$.

In theoretical side, the aTGC such as the anomalous $W^+W^-V$ $\left(V=\gamma,Z\right)$ couplings have been discussed previously in the literature \cite{Baur:1988tfa,Hagiwara:1987xdj,Hagiwara:1992ghg,Wiest:1995yjk,Gintner:1995kkd,Ambrosanio:1992lkj,Sahin:2011dfg,Kepka:2008vmn,Ari:2016aac,Atag:2001ata,Koksal:2019opr,Rodriguez:2019erv,Spor:2020ghy,Sahin:2017uot,Bian:2016wer,Choudhury:1997xsa,Choudhury:1997mep,Li:2018tyb,Kumar:2015ghe,Falkowski:2015ghe,Bhatia:2019gso,Etesami:2016eto,Cakir:2014mjp,Sahin:2009tbz,Atag:2001mai,Bian:2015ylk,Rahaman:2020abs,Baer:2013ttt,Billur:2019suu}. The anomalous $W^+W^-\gamma$ couplings have been studied experimentally on the parameters of $\kappa_{\gamma}$ and $\lambda_{\gamma}$ at the LEP \cite{Schael:2013msh,Abbiendi:2001yuo,Abdallah:2008bnw}, the Tevatron \cite{Aaltonen:2007uas,Aaltonen:2009skw,Aaltonen:2010tyf,Abazov:2012mwc} and the LHC \cite{Chatrchyan:2013ewm,Aaboud:2017les,Sirunyan:2017txe,Sirunyan:2019umc}. The limits of the anomalous coupling parameters on the aTGC obtained in some experimental and the phenomenological studies are given in Table~\ref{tab1}.

\begin{table}
\caption{The best limits at 95$\%$ C.L. on the aTGC with $\Delta \kappa_{\gamma}$, $\lambda_{\gamma}$, $c_{WWW}/\Lambda^2$, $c_{W}/\Lambda^2$ and $c_{B}/\Lambda^2$ parameters obtained from the experimental and the phenomenological studies.}
\label{tab1}
\begin{ruledtabular}
\begin{tabular}{cccccc}
\multirow{2}{*}{Experimental limit} & \multirow{2}{*}{$\Delta \kappa_{\gamma}$} & \multirow{2}{*}{$\lambda_{\gamma}$} & $c_{WWW}/\Lambda^2$ & $c_{W}/\Lambda^2$ & $c_{B}/\Lambda^2$\\ 
& & & (TeV$^{-2}$) & (TeV$^{-2}$) & (TeV$^{-2}$)\\
\hline
CMS & \multirow{2}{*}{[-0.0275; 0.0286]} & \multirow{2}{*}{[-0.0065; 0.0066]} & \multirow{2}{*}{[-1.58; 1.59]} & \multirow{2}{*}{[-2.00; 2.65]} & \multirow{2}{*}{[-8.78; 8.54]}\\
Collaboration \cite{Sirunyan:2019umc} & & & & & \\ \hline
ATLAS & \multirow{2}{*}{[-0.0610; 0.0640]} & \multirow{2}{*}{[-0.0130; 0.0130]} & \multirow{2}{*}{[-3.10; 3.10]} & \multirow{2}{*}{[-5.10; 5.80]} & \multirow{2}{*}{[-19.0; 20.0]}\\
Collaboration \cite{Aaboud:2017les} & & & & & \\ \hline
CDF & \multirow{2}{*}{[-0.5700; 0.6500]} & \multirow{2}{*}{[-0.1400; 0.1500]} & \multirow{2}{*}{[-34.1; 36.0]} & \multirow{2}{*}{[-53.0; 72.3]} & \multirow{2}{*}{[-166; 178]}\\
Collaboration \cite{Aaltonen:2010tyf} & & & & & \\ \hline
D0 & \multirow{2}{*}{[-0.1580; 0.2550]} & \multirow{2}{*}{[-0.0360; 0.0440]} & \multirow{2}{*}{[-8.70; 11.0]} & \multirow{2}{*}{[-8.20; 20.0]} & \multirow{2}{*}{[-53.0; 78.1]}\\
Collaboration \cite{Abazov:2012mwc} & & & & & \\ \hline
ALEP, DELPHI, & \multirow{2}{*}{[-0.0990; 0.0660]} & \multirow{2}{*}{[-0.0590; 0.0170]} & \multirow{2}{*}{[-14.0; 4.10]} & \multirow{2}{*}{[-13.0; 5.10]} & \multirow{2}{*}{[-25.1; 20.1]}\\
L3, OPAL \cite{Schael:2013msh} & & & & & \\ \hline \hline
Phenomenological & \multirow{2}{*}{$\Delta \kappa_{\gamma}$} & \multirow{2}{*}{$\lambda_{\gamma}$} & & & \\ 
limit & & & & & \\
\hline
ILC \cite{Rahaman:2020abs} & [-0.0021; 0.0015] & [-0.0021; 0.0018] & & & \\
ILC \cite{Baer:2013ttt} & [-0.00037; 0.00037] & [-0.00051; 0.00051] &  &  & \\
CLIC \cite{Billur:2019suu} & [-0.00007; 0.00007] & [-0.00004; 0.00102] &  &  & \\
CLIC \cite{Ari:2016aac} & [-0.0004; 0.0023] & [-0.0007; 0.0007] &  &  & \\
CEPC \cite{Bian:2015ylk} & [-0.00045; 0.00045] & [-0.00033; 0.00033] &  &  & \\
CEPC \cite{Ari:2016aac} & [-0.00102; 0.00103] & [-0.00168; 0.00173] &  &  & \\
\end{tabular}
\end{ruledtabular}
\end{table}

\section{Future lepton collider: ILC}

The SM is a successful theory that answers many important questions in particle physics, such as describing electromagnetic, weak and strong interactions in the universe and predicting all known elementary particles. Although the SM has passed all experimental tests, some significant arguments such as the hierarchy problem, the non-unification of fundamental forces, the baryon-antibaryon asymmetry, non-explained dark matter demonstrate that the SM has some shortcomings to be final theory of everything. For this reason, there is a great desire to search for the new physics beyond the SM.

The colliders in experimental particle physics are often classified according to their shape (linear/circular) and the type of colliding particles (hadron/lepton). All major differences between the hadron and lepton colliders depend on the nature of the colliding particles. The leptons at the lepton collider are elementary particles. Because the initial state of elementary particles is fully defined at the fundamental level, the collisions in this collider are clean without hadronic activity and the measurements are precise. The hadrons at the hadron collider are composite particles, composed of quarks, which are elementary particles. Since hadrons are heavier than leptons, the hadron colliders have higher collision energy than the lepton colliders. Although high collision energies play a key role in investigating the new particles and their interactions, each collision in the hadron collider composes the backgrounds for physics analysis, creating a large number of elementary processes. These leaves high doses of radiation energy to the detector and thus it becomes difficult to perform analysis and make precision measurements.

The discovery potential of the LHC would be complemented by the ILC, which is a linear electron-positron collider in the design phase \cite{Bambade:2019pkd}. The ILC is planned to reach tunable center-of-mass energy up to 500 GeV (upgradeable up to 1 TeV) with published the Technical Design Report (TDR) for the ILC accelerator \cite{Adolphsen:2013kgd,Adolphsen:2013rew}. The electron and the positron beams are longitudinally polarized to $80\%$ and $30\%$, respectively, which have different signs of polarization from each other. The positron beams are foreseen to polarize about $60\%$ in the upgraded option of the ILC. While polarized electron beams are produced by photoproduction with a polarized laser, polarized positron beams are produced in pair conversion $\gamma\,\rightarrow\,e^+e^-$, where the photon is produced by a high energy electron beam passing through a superconducting undulator \cite{Potter:2020ghb}. A Compton polarimeter will perform the primary polarimeter measurement at the ILC. Polarimeters can be placed upstream or downstream of the Interference Point (IP). Because of distortion and radiation in the beam-beam collision process luminosity-weighted beam polarization will differ from measured polarization. The difference between  luminosity-weighted beam polarization and polarimeter measurement is expressed as $dP=P_z^{\text{lum-wt}}-P_z^{\text{CIP}}$. The beam direction at the IP of the polarimeter (Compton IP) needs to be aligned within 50 $\text{$\mu$}$rad with the collision axis at $e^+e^-$ IP for achieve the minimization of $dP$ \cite{Moortgat:2008yhn}. Using both upstream and downstream polarimetry will provide to obtain the desired minimum $dP$ and predict the systematic error.

Thanks to the clean event environment, the tunable collision energy and the potential to polarize beams, it is possible for the ILC to observe the smallest deviation from SM predictions indicating new physics as well as to discover new particles and to make their precise measurements \cite{Moortgat:2008yhn}. The possibility of both electron and positron beam polarization in the ILC is of great importance in reaching the major goals of particle physics. The two polarized beams in the ILC are very powerful tools to reveal the structure of the underlying physics, determine new physics parameters in model-independent analysis and also test basic model assumptions. 

It is also useful to compare the effects of unpolarized and polarized positron beams. Because the positron polarization would dramatically improve the search potential for new particles, the disentanglement of their dynamics and the accuracy at which left-right asymmetry observables can be measured. The positron polarization option greatly reduces model dependency, allowing more observables to be measured. If the main benefits of positron polarization can be divided into three parts \cite{Moortgat:2008yhn}, firstly, positron polarization enables us to acquire subsamples of the data with higher rates for physics processes and lower rates for backgrounds. Secondly, positron polarization presents four distinct data sets instead of the two available if only the electron beam can be polarized. The flexibility in choosing between different data set is a very important feature of the ILC. Finally, the likely most striking effect is the control of systematic uncertainties. Here, the precisions targeting the ILC are achieved when all systematic uncertainties are controlled to the same level or better as statistical uncertainties.

Proper beam polarizations combined with high luminosity increase analysis capability, resulting in better statistics and reduction of systematic errors, and enable to enhance or suppress the SM processes and to reveal new processes. Hence, the proper combinations of the polarized electron and positron beams are useful in increasing signal rates and minimizing unwanted background processes. Each specified capability of the ILC provides a great number of additional opportunities.

The cross section of any process is determined from the four possible pure chiral cross sections with electron beam polarizations $P_{e^-}$ and positron beam polarization $P_{e^+}$ by \cite{Fujii:2018tgt}

\begin{eqnarray}
\label{eq.14} 
\sigma\left(P_{e^-},P_{e^+}\right)=\frac{1}{4}\{\left(1+P_{e^-}\right)\left(1+P_{e^+}\right)\sigma_{RR}+\left(1-P_{e^-}\right)\left(1-P_{e^+}\right)\sigma_{LL}
\\ \nonumber
+\left(1+P_{e^-}\right)\left(1-P_{e^+}\right)\sigma_{RL}+\left(1-P_{e^-}\right)\left(1+P_{e^+}\right)\sigma_{LR}\}\,,
\end{eqnarray}

where $\sigma_{LR}$ represents the cross section if the electron beam is left-handed polarized ($P_{e^-}=-1$) and the positron beam is right-handed polarized ($P_{e^+}=+1$). Other cross sections are similarly defined. The unpolarized cross section $\sigma_0$ is expressed as

\begin{eqnarray}
\label{eq.15} 
\sigma_0=\frac{1}{4}\{\sigma_{RR}+\sigma_{LL}+\sigma_{RL}+\sigma_{LR}\}\,.
\end{eqnarray}

Also, the other significant definitions are the effective polarization and the effective luminosity \cite{Omori:1999gek,Renard:1989thq,Jacob:1959rer,Jacob:2000fty,Renard:1981tda}. The formulas for these definitions are given below:

\begin{eqnarray}
\label{eq.16} 
{\cal L}_{\text{eff}}=\frac{1}{2}\left(1-P_{e^-}P_{e^+}\right){\cal L}\,,
\end{eqnarray}
\begin{eqnarray}
\label{eq.17} 
P_{\text{eff}}=\frac{P_{e^-}-P_{e^+}}{1-P_{e^-}P_{e^+}}\,.
\end{eqnarray}

The effective polarization and the ratio ${\cal L}_{\text{eff}}/{\cal L}$ according to some electron and positron beam polarizations are obtained from Eqs.~(\ref{eq.16})-(\ref{eq.17}) and given in Table~\ref{tab2}. Since beam polarization plays a significant role by effectively enhancing the signal and suppressing the background rates, a scaling factor comparing the cross-sections with two different polarization configurations is parametrized by:

\begin{eqnarray}
\label{eq.18} 
\text{scaling factor}=\frac{\sigma^{\left(P_{e^-},P_{e^+}\right)_{\left(b\right)}}}{\sigma^{\left(P_{e^-},P_{e^+}\right)_{\left(a\right)}}}\,,
\end{eqnarray}

where for two different cross sections, the one with the index $\left(a\right)$ indicates that only the electron beam is polarized and the one with the index $\left(b\right)$ indicates that both beams are polarized. A scaling factor can range from 0 to maximum of 2. For example, using the results at Table~\ref{tab4} in Section III of this paper, it is seen that the cross sections for $c_{WWW}/\Lambda^2$ coupling and Cut-3d are $29.27$ pb at $\left(P_{e^-},P_{e^+}\right)=\left(+80\%,0\%\right)$ polarization, $20.48$ pb at $\left(P_{e^-},P_{e^+}\right)=\left(+80\%,-30\%\right)$ polarization and $11.78$ pb at $\left(P_{e^-},P_{e^+}\right)=\left(+80\%,-60\%\right)$ polarization. As a result, the scaling factors are calculated from Eq.~(\ref{eq.18}) to about $0.7$ between $\left(P_{e^-},P_{e^+}\right)=\left(+80\%,-30\%\right)$ and $\left(P_{e^-},P_{e^+}\right)=\left(+80\%,0\%\right)$ polarizations and about $0.4$ between $\left(P_{e^-},P_{e^+}\right)=\left(+80\%,-60\%\right)$ and $\left(P_{e^-},P_{e^+}\right)=\left(+80\%,0\%\right)$ polarizations.

\begin{table}[H]
\caption{Effective polarization and effective luminosity for electron and positron beam polarizations.}
\label{tab2}
\centering
\begin{tabular}{ P{3cm} P{3cm} P{3cm} P{3cm} }
\hline \hline
$P_{e^-}$ & $P_{e^+}$ & $P_{\text{eff}}$ & ${\cal L}_{\text{eff}}/{\cal L}$ \\ 
\hline
$-100\%$ & $0$ & $-100\%$ & $0.50$ \\ 
$-80\%$ & $0$ & $-80\%$ & $0.50$ \\ 
$-80\%$ & $+30\%$ & $-89\%$ & $0.62$ \\ 
$-80\%$ & $+60\%$ & $-95\%$ & $0.74$ \\ \hline \hline
\end{tabular}
\end{table}

\section{Cross sections and sensitivity analysis of the process $e^-e^+\,\rightarrow\,\nu_e\overline{\nu}_e\gamma$ at the ILC}

\begin{figure}[H]
\centering
\includegraphics[scale=0.61]{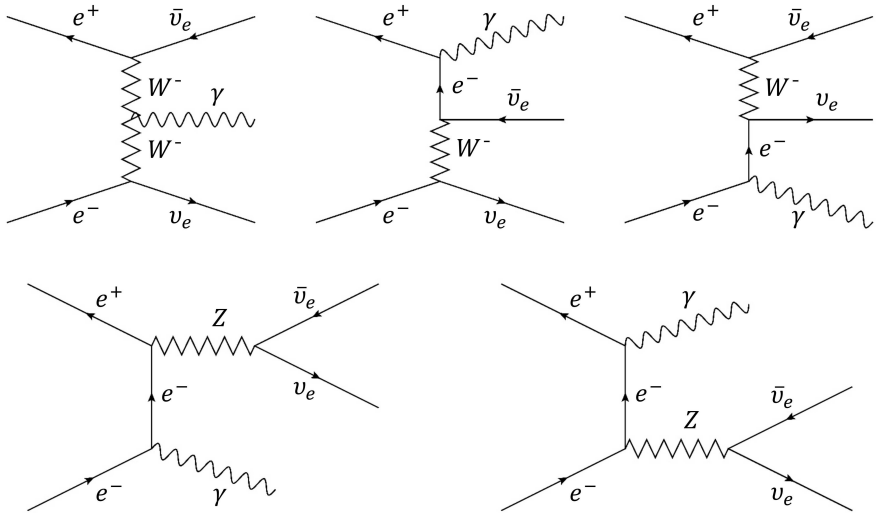}
\caption{The Feynman diagrams for the process $e^- e^+ \,\rightarrow\,\nu_e \overline{\nu}_e \gamma$. 
\label{fig1}}
\end{figure}

The Feynman diagrams for the process $e^-e^+\,\rightarrow\,\nu_e\overline{\nu}_e\gamma$ are given in Fig.~\ref{fig1}. The first of the five Feynman diagrams includes the anomalous $W^+W^-\gamma$ coupling and it contributes to the new physics. We can see from Fig.~\ref{fig1} that one of the advantages of the process $e^- e^+ \,\rightarrow\,\nu_e \overline{\nu}_e \gamma$ is that they can isolate $W^+W^-\gamma$ couplings from $W^+W^-Z$ couplings. Also, the neutrinos are of major importance to the elementary particle theory, astrophysics and cosmology \cite{Fargion:1996fet}. In our calculations, the total cross sections of the process $e^-e^+\,\rightarrow\,\nu_e\overline{\nu}_e\gamma$ with the configurations of electron-positron beam polarization are simulated using {\sc MadGraph5} aMC@NLO \cite{Alwall:2014cvc}. At this program, we use the EWdim6 model file for the operators that examine interactions between the electroweak gauge boson we have described above in dimension-six. We examine the potential of the process $e^-e^+\,\rightarrow\,\nu_e\overline{\nu}_e\gamma$ at the ILC with $\sqrt{s}=500$ GeV. The configurations of electron-positron beam polarization and their corresponding integrated luminosities are studied at the ILC:

\begin{eqnarray}
\label{eq.19} 
\left(P_{e^-},P_{e^+}\right)=\left(\mp80\%,\pm60\%\right)\, \text{and}\,\, {\cal L}=1350\,\, \text{fb}^{-1}\,,
\nonumber \\
\left(P_{e^-},P_{e^+}\right)=\left(\mp80\%,\pm30\%\right)\, \text{and}\,\, {\cal L}=1600\,\, \text{fb}^{-1}\,, 
\nonumber \\
\left(P_{e^-},P_{e^+}\right)=\left(\pm80\%,0\%\right)\, \text{and}\,\, {\cal L}=2000\,\, \text{fb}^{-1}\,,
\\
\text{unpolarized electron-positron beam and}\,\, {\cal L}=4000\,\, \text{fb}^{-1}\,. \nonumber 
\end{eqnarray}

{\raggedright Here, we have used $-$ sign for left polarization and $+$ sign for right polarization.}

We apply the kinematic selection cuts to suppress the backgrounds and to optimize the signal sensitivity. $p_T^\nu$ is transverse momentum of the final state neutrinos, $\left|\eta^\gamma\right|$ is the pseudorapidity of the photon and $p_T^\gamma$ is the transverse momentum of the photon. The outgoing particles are required to satisfy these kinematic cuts for $\nu_e\overline{\nu}_e\gamma$ events at the ILC. We consider $p_T^\nu>25$ GeV with tagged Cut-1, $\left|\eta^\gamma\right|<2.5$ with tagged Cut-2 and four different values of the transverse momentum of the photon, $p_T^\gamma>10,15,20,25$ GeV with tagged Cut-3a, Cut-3b, Cut-3c and Cut-3d, respectively. A summary of the kinematic cuts is given in Table~\ref{tab3}.

\begin{table}[H]
\centering
\caption{Definitions of kinematic cuts used for the analysis.}
\label{tab3}
\begin{tabular}{p{3cm}p{4cm}}
\hline \hline
Cuts & Definitions \\ 
\hline
Cut-1 & $p_T^\nu >25$ GeV\\ 
Cut-2 & Cut-1 $+$ $|\eta^\gamma|<2.5$\\ 
Cut-3a & Cut-2 $+$ $p_T^\gamma >10$ GeV\\
Cut-3b & Cut-2 $+$ $p_T^\gamma >15$ GeV\\
Cut-3c & Cut-2 $+$ $p_T^\gamma >20$ GeV\\
Cut-3d & Cut-2 $+$ $p_T^\gamma >25$ GeV\\ \hline \hline
\end{tabular}
\end{table}

\begin{figure}[H]
\includegraphics[scale=0.61]{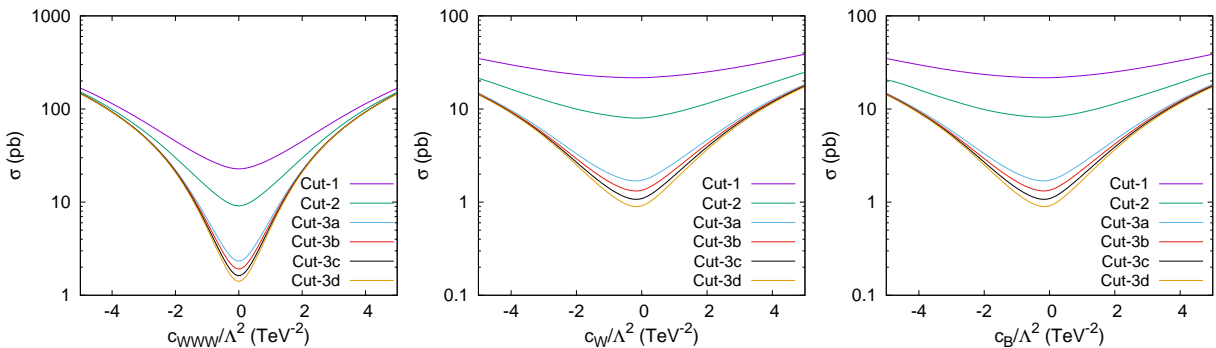}
\caption{The total cross sections of the main process $e^- e^+ \,\rightarrow\,\nu_e \overline{\nu}_e \gamma$ for unpolarized electron-positron beam as a function of $c_{WWW}/\Lambda^2$, $c_{W}/\Lambda^2$ and $c_{B}/\Lambda^2$ at six different kinematic cuts. 
\label{fig2}}
\end{figure}

\begin{figure}[H]
\includegraphics[scale=0.61]{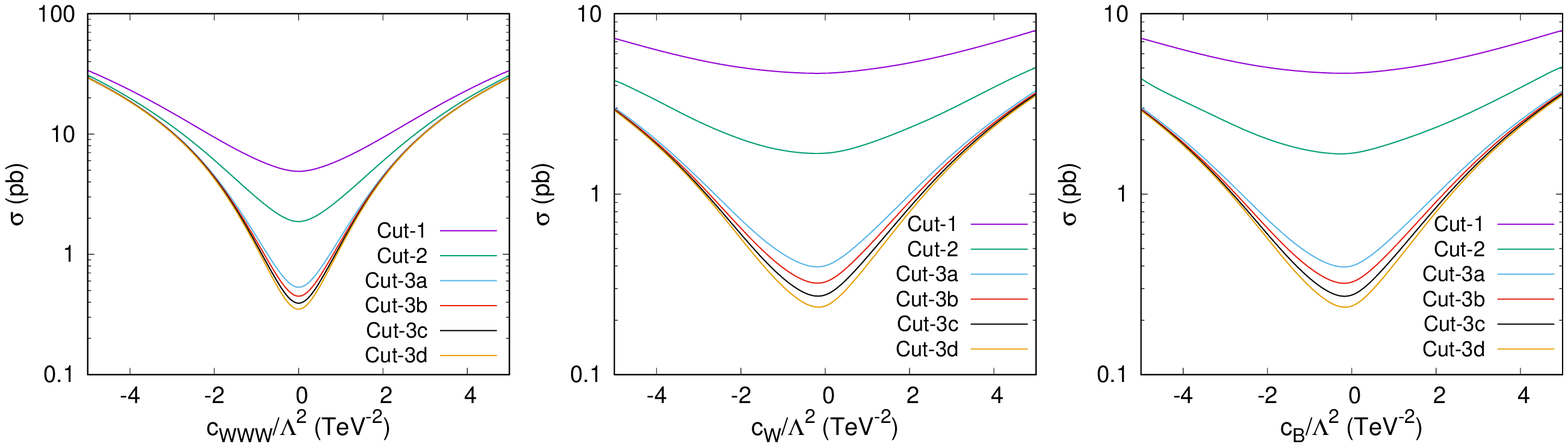}
\caption{Same as in Fig.~\ref{fig2}, but for $\left(P_{e^-},P_{e^+}\right)=\left(+80\%,0\%\right)$.     
\label{fig3}}
\end{figure}

\begin{figure}[H]
\includegraphics[scale=0.61]{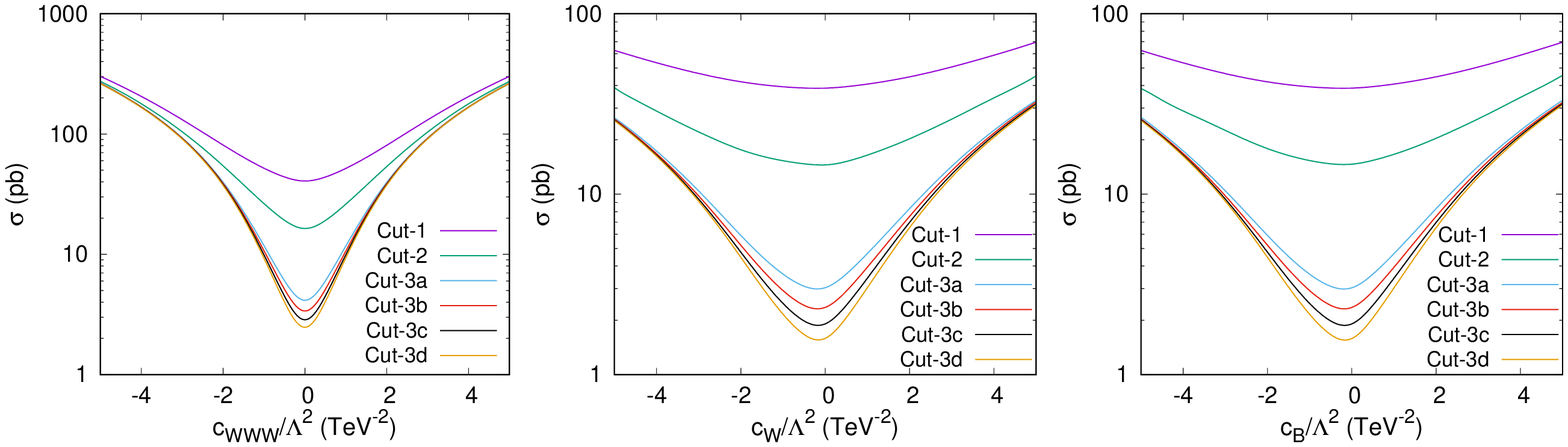}
\caption{Same as in Fig.~\ref{fig2}, but for $\left(P_{e^-},P_{e^+}\right)=\left(-80\%,0\%\right)$.   
\label{fig4}}
\end{figure}

\begin{figure}[H]
\includegraphics[scale=0.61]{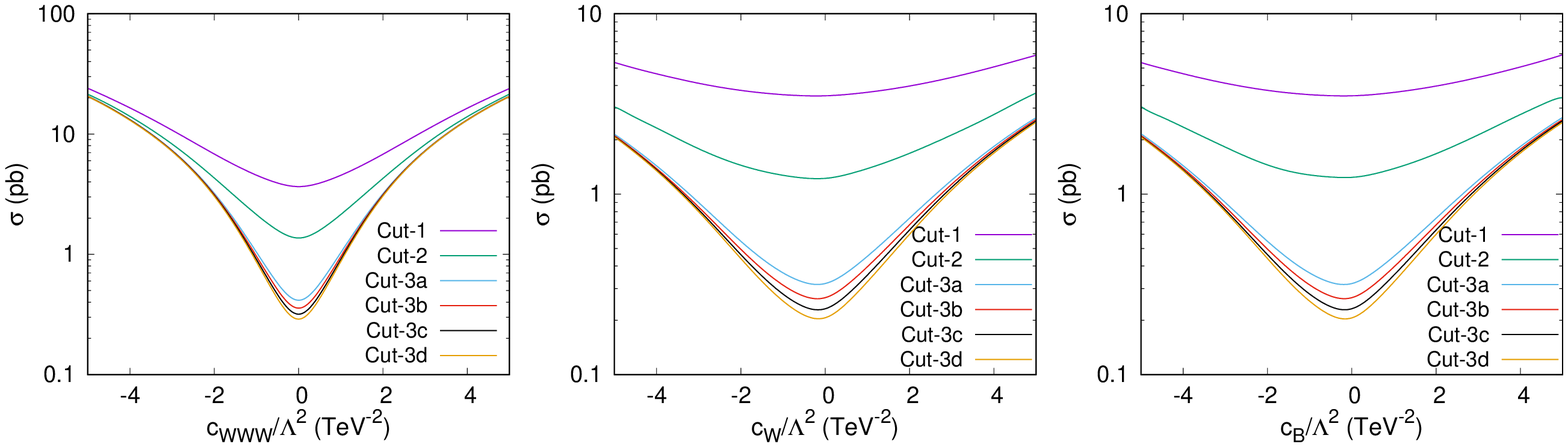}
\caption{Same as in Fig.~\ref{fig2}, but for $\left(P_{e^-},P_{e^+}\right)=\left(+80\%,-30\%\right)$.     
\label{fig5}}
\end{figure}

\begin{figure}[H]
\includegraphics[scale=0.61]{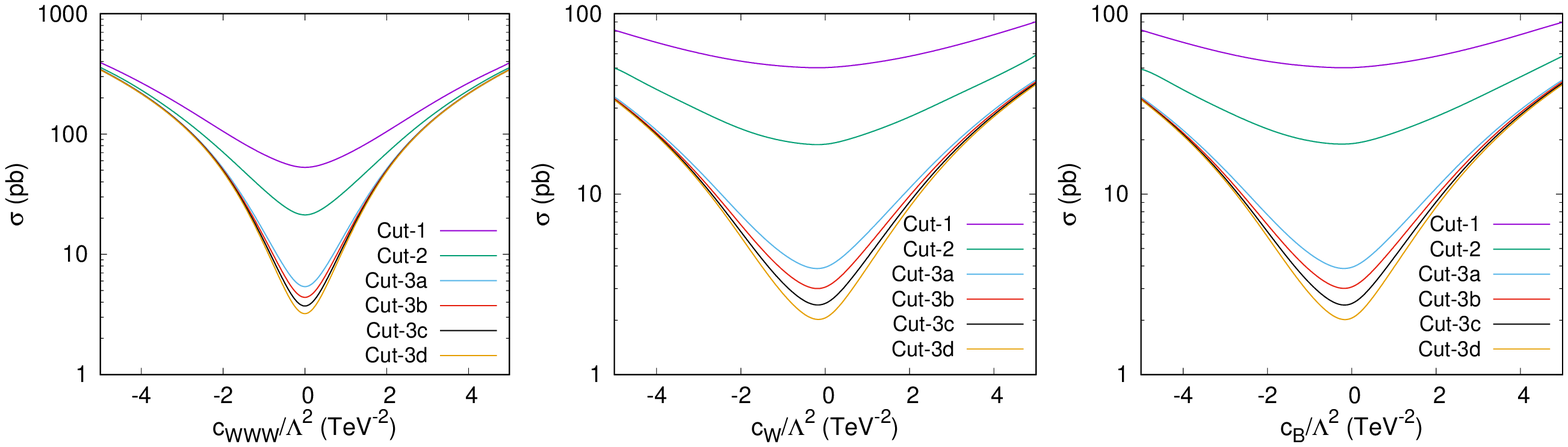}
\caption{Same as in Fig.~\ref{fig2}, but for $\left(P_{e^-},P_{e^+}\right)=\left(-80\%,+30\%\right)$.      
\label{fig6}}
\end{figure}

\begin{figure}[H]
\includegraphics[scale=0.61]{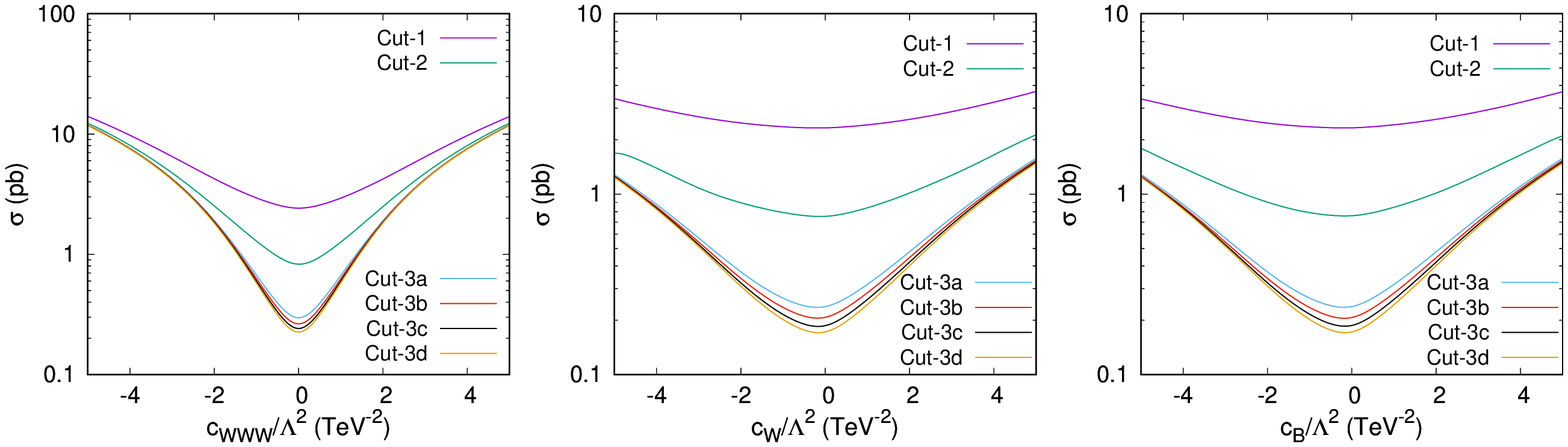}
\caption{Same as in Fig.~\ref{fig2}, but for $\left(P_{e^-},P_{e^+}\right)=\left(+80\%,-60\%\right)$.     
\label{fig7}}
\end{figure}

\begin{figure}[H]
\includegraphics[scale=0.61]{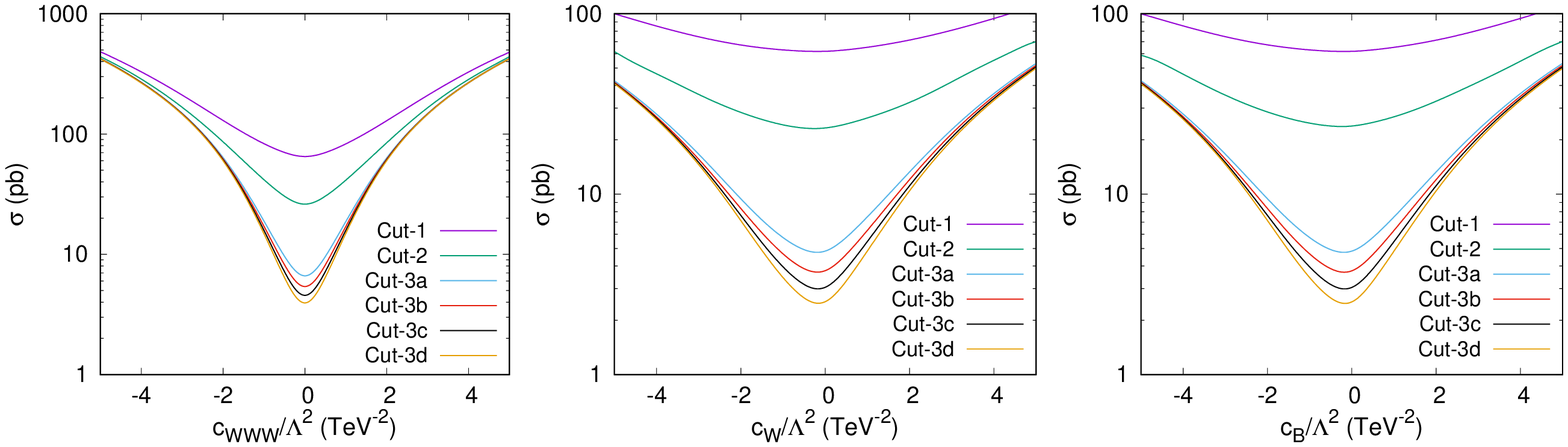}
\caption{Same as in Fig.~\ref{fig2}, but for $\left(P_{e^-},P_{e^+}\right)=\left(-80\%,+60\%\right)$.      
\label{fig8}}
\end{figure}

In this analysis, we focus on CP-conserving $c_{WWW}/\Lambda^2$, $c_{W}/\Lambda^2$ and $c_{B}/\Lambda^2$ couplings via the process $e^-e^+\,\rightarrow\,\nu_e\overline{\nu}_e\gamma$ at the ILC. The total cross sections of the process $e^-e^+\,\rightarrow\,\nu_e\overline{\nu}_e\gamma$ as a function of anomalous $c_{WWW}/\Lambda^2$, $c_{W}/\Lambda^2$ and $c_{B}/\Lambda^2$ couplings parameters for kinematic cuts defined in Table~\ref{tab3} are presented in Fig.~\ref{fig2} with unpolarized electron-positron beam, in Fig.~\ref{fig3} with polarized electron-positron beam $\left(P_{e^-},P_{e^+}\right)=\left(+80\%,0\%\right)$, in Fig.~\ref{fig4} with polarized electron-positron beam $\left(P_{e^-},P_{e^+}\right)=\left(-80\%,0\%\right)$, in Fig.~\ref{fig5} with polarized electron-positron beam $\left(P_{e^-},P_{e^+}\right)=\left(+80\%,-30\%\right)$, in Fig.~\ref{fig6} with polarized electron-positron beam $\left(P_{e^-},P_{e^+}\right)=\left(-80\%,+30\%\right)$, in Fig.~\ref{fig7} with polarized electron-positron beam $\left(P_{e^-},P_{e^+}\right)=\left(+80\%,-60\%\right)$ and in Fig.~\ref{fig8} with polarized electron-positron beam $\left(P_{e^-},P_{e^+}\right)=\left(-80\%,+60\%\right)$. This total cross sections are calculated with center-of-mass energy of $\sqrt{s}=500$ GeV at the ILC and depend on the integrated luminosities corresponding to configurations of electron-positron beam polarization in Eq.~(\ref{eq.19}). While the cross section curves belonging to the $c_{W}/\Lambda^2$ and $c_{B}/\Lambda^2$ couplings in Figs.~\ref{fig2}-\ref{fig8} show a similar characteristic due to Eq.~(\ref{eq.12}), the cross section curves belonging to the $c_{WWW}/\Lambda^2$ coupling differ from that of the $c_{W}/\Lambda^2$ and $c_{B}/\Lambda^2$ couplings. Also, in Figs.~\ref{fig2}-\ref{fig8}, the total cross section values as a function of the $c_{WWW}/\Lambda^2$ coupling have larger than that as a function of the $c_{W}/\Lambda^2$ and $c_{B}/\Lambda^2$ couplings. The suppression in the cutflow at Table~\ref{tab3} increases from Cut-1 to Cut-3d and this increment cause a decrease in the total cross sections from Cut-1 to Cut-3d in Figs.~\ref{fig2}-\ref{fig8}.

We present the total and SM cross sections of the process $e^-e^+\,\rightarrow\,\nu_e\overline{\nu}_e\gamma$ for the kinematic cuts of seven polarization scenarios with respect to $c_{WWW}/\Lambda^2$, $c_{W}/\Lambda^2$ and $c_{B}/\Lambda^2$ couplings in Table~\ref{tab4}. The total and SM cross section values for $c_{WWW}/\Lambda^2$, $c_{W}/\Lambda^2$ and $c_{B}/\Lambda^2$ couplings in Table~\ref{tab4} correspond to $c_{WWW}/\Lambda^2$, $c_{W}/\Lambda^2$ and $c_{B}/\Lambda^2=5$ TeV$^{-2}$, respectively. The ratios arise from the total cross sections divided by the SM cross sections and increase after each applied kinematic cut. As the applied kinematic cuts increase, the SM cross section is suppressed, thus the signal becomes more apparent.

\begin{table}[H]
\caption{Total and SM cross section values and the ratios of total cross section to SM cross section on the anomalous $c_{WWW}/\Lambda^2$, $c_{W}/\Lambda^2$ and $c_{B}/\Lambda^2$ couplings for seven different polarization scenarios and six different cuts.}
\label{tab4}
\begin{ruledtabular}
\begin{tabular}{ccccccccc}
\multicolumn{3}{c}{} & \multicolumn{2}{c}{$c_{WWW}/\Lambda^2$ (TeV$^{-2}$)} & \multicolumn{2}{c}{$c_{W}/\Lambda^2$ (TeV$^{-2}$)} & \multicolumn{2}{c}{$c_{B}/\Lambda^2$ (TeV$^{-2}$)}\\
\hline
Polarization & \multirow{2}{*}{Cuts} & SM cross & Total cross & \multirow{2}{*}{Ratio} & Total cross & \multirow{2}{*}{Ratio} & Total cross & \multirow{2}{*}{Ratio} \\ 
scenarios &  & sections & sections &  & sections &  & sections &  \\ 
 &  & (pb) & (pb) &  & (pb) &  & (pb) &  \\
\hline \hline
\multirow{6}{*}{Unpolarized} 
 & Cut-1 & 21.53 & 167.04 & 7.75 & 38.72 & 1.79 & 38.75 & 1.79\\
 & Cut-2 & 8.07 & 151.64 & 18.79 & 24.68 & 3.05 & 24.39 & 3.02\\ 
 & Cut-3a & 1.58 & 146.89 & 92.96 & 18.40 & 11.64 & 18.45 & 11.67\\
 & Cut-3b & 1.23 & 146.43 & 119.04 & 18.00 & 14.63 & 17.94 & 14.58\\
 & Cut-3c & 0.98 & 146.17 & 149.15 & 17.64 & 18.00 & 17.60 & 17.95\\
 & Cut-3d & 0.81 & 145.70 & 179.87 & 17.36 & 21.43 & 17.31 & 21.37\\ \hline

 & Cut-1 & 4.66 & 33.71 & 7.23 & 8.09 & 1.73 & 8.07 & 1.73\\ 
 & Cut-2 & 1.72 & 30.80 & 17.90 & 5.03 & 2.92 & 5.07 & 2.94\\ 
$P_{e^-}=+80\%;$ & Cut-3a & 0.37 & 29.47 & 79.64 & 3.74 & 10.10 & 3.73 & 10.08\\
$P_{e^+}=0$ & Cut-3b & 0.30 & 29.39 & 97.96 & 3.64 & 12.13 & 3.65 & 12.16\\
 & Cut-3c & 0.25 & 29.34 & 117.36 & 3.58 & 14.32 & 3.60 & 14.40\\
 & Cut-3d & 0.21 & 29.27 & 139.38 & 3.51 & 16.71 & 3.52 & 16.76\\ \hline

 & Cut-1 & 38.44 & 301.79 & 7.85 & 69.43 & 1.80 & 69.53 & 1.80\\ 
 & Cut-2 & 15.28 & 273.91 & 17.92 & 45.24 & 2.96 & 45.51 & 2.97\\ 
$P_{e^-}=-80\%;$ & Cut-3a & 2.81 & 264.55 & 94.14 & 33.09 & 11.77 & 33.12 & 11.78\\
$P_{e^+}=0$ & Cut-3b & 2.15 & 263.55 & 122.58 & 32.47 & 15.10 & 32.26 & 15.00\\
 & Cut-3c & 1.71 & 263.16 & 153.89 & 31.68 & 18.52 & 31.65 & 18.50\\
 & Cut-3d & 1.40 & 263.07 & 187.90 & 31.15 & 22.25 & 31.19 & 22.27\\ \hline

Continued on next page
\end{tabular}
\end{ruledtabular}
\end{table}

\begin{table}[H]
\renewcommand\thetable{IV}
\caption{Continued from previous page.}
\label{tab4}
\begin{ruledtabular}
\begin{tabular}{ccccccccc}
\multicolumn{3}{c}{} & \multicolumn{2}{c}{$c_{WWW}/\Lambda^2$ (TeV$^{-2}$)} & \multicolumn{2}{c}{$c_{W}/\Lambda^2$ (TeV$^{-2}$)} & \multicolumn{2}{c}{$c_{B}/\Lambda^2$ (TeV$^{-2}$)}\\
\hline
Polarization & \multirow{2}{*}{Cuts} & SM cross & Total cross & \multirow{2}{*}{Ratio} & Total cross & \multirow{2}{*}{Ratio} & Total cross & \multirow{2}{*}{Ratio} \\ 
scenarios &  & sections & sections &  & sections &  & sections &  \\ 
 &  & (pb) & (pb) &  & (pb) &  & (pb) &  \\
\hline \hline

 & Cut-1 & 3.46 & 23.87 & 6.89 & 5.89 & 1.70 & 5.92 & 1.71\\ 
 & Cut-2 & 1.27 & 21.48 & 16.91 & 3.63 & 2.85 & 3.42 & 2.69\\ 
$P_{e^-}=+80\%;$ & Cut-3a & 0.30 & 20.60 & 68.66 & 2.65 & 8.83 & 2.66 & 8.86\\
$P_{e^+}=-30\%$ & Cut-3b & 0.25 & 20.57 & 82.28 & 2.59 & 10.36 & 2.59 & 10.36\\
 & Cut-3c & 0.21 & 20.51 & 97.66 & 2.54 & 12.09 & 2.54 & 12.09\\
 & Cut-3d & 0.18 & 20.48 & 113.77 & 2.50 & 13.88 & 2.50 & 13.88\\ \hline
 
 & Cut-1 & 50.01 & 390.60 & 7.81 & 90.31 & 1.80 & 89.75 & 1.79\\ 
 & Cut-2 & 18.47 & 355.91 & 19.26 & 59.14 & 3.20 & 58.26 & 3.15\\ 
$P_{e^-}=-80\%;$ & Cut-3a & 3.64 & 344.27 & 94.57 & 43.04 & 11.82 & 42.96 & 11.80\\
$P_{e^+}=+30\%$ & Cut-3b & 2.79 & 342.66 & 122.81 & 42.06 & 15.07 & 42.09 & 15.08\\
 & Cut-3c & 2.23 & 342.20 & 153.45 & 41.24 & 18.49 & 41.25 & 18.49\\
 & Cut-3d & 1.82 & 341.61 & 187.69 & 40.56 & 22.28 & 40.54 & 22.27\\ \hline 
 
 & Cut-1 & 2.31 & 13.97 & 6.04 & 3.71 & 1.60 & 3.69 & 1.59\\ 
 & Cut-2 & 0.70 & 12.26 & 17.51 & 2.13 & 3.04 & 2.10 & 3.00\\ 
$P_{e^-}=+80\%;$ & Cut-3a & 0.22 & 11.82 & 53.72 & 1.58 & 7.18 & 1.57 & 7.13\\
$P_{e^+}=-60\%$ & Cut-3b & 0.19 & 11.81 & 62.15 & 1.53 & 8.05 & 1.53 & 8.05\\
 & Cut-3c & 0.17 & 11.80 & 69.41 & 1.51 & 8.88 & 1.50 & 8.82\\
 & Cut-3d & 0.16 & 11.78 & 73.62 & 1.48 & 9.25 & 1.48 & 9.25\\ \hline
 
 & Cut-1 & 61.42 & 479.29 & 7.80 & 111.24 & 1.81 & 111.14 & 1.80\\ 
 & Cut-2 & 23.46 & 439.17 & 18.71 & 69.86 & 2.97 & 70.20 & 2.99\\ 
$P_{e^-}=-80\%;$ & Cut-3a & 4.51 & 423.21 & 93.83 & 52.97 & 11.74 & 53.05 & 11.76\\
$P_{e^+}=+60\%$ & Cut-3b & 3.44 & 422.43 & 122.79 & 51.73 & 15.03 & 51.78 & 15.05\\
 & Cut-3c & 2.76 & 421.42 & 152.68 & 50.71 & 18.37 & 50.74 & 18.38\\
 & Cut-3d & 2.24 & 420.53 & 187.73 & 49.88 & 22.26 & 49.76 & 22.21\\ 
\end{tabular}
\end{ruledtabular}
\end{table}

\begin{figure}[H]
\includegraphics[scale=0.61]{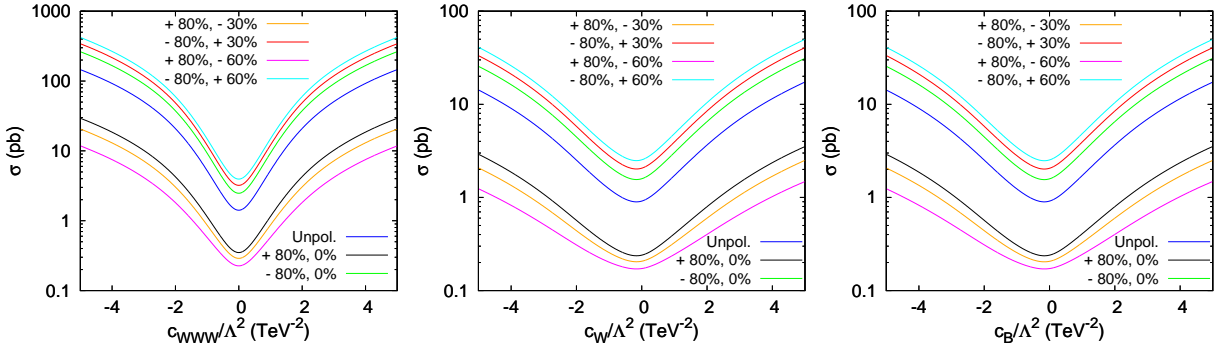}
\caption{The total cross sections of the main process $e^- e^+ \,\rightarrow\,\nu_e \overline{\nu}_e \gamma$ as a function of $c_{WWW}/\Lambda^2$, $c_{W}/\Lambda^2$ and $c_{B}/\Lambda^2$ for seven different polarization scenarios at Cut-3d. 
\label{fig9}}
\end{figure}

The total cross sections of the process $e^- e^+ \,\rightarrow\,\nu_e \overline{\nu}_e \gamma$ as a function of $c_{WWW}/\Lambda^2$, $c_{W}/\Lambda^2$ and $c_{B}/\Lambda^2$ at the largest cut (Cut-3d) according to the polarization configurations are compared in Fig.~\ref{fig9} and thus the effect of polarizations on the total cross sections is observed. The left-polarized electron (right-polarized positron) beam enhances the cross sections due to the structure of the $e^-\nu_eW^-$ $\left(e^+\overline{\nu}_eW^-\right)$ vertex in the first Feynmann diagram of Fig.~\ref{fig1}, which contains the largest contribution with the anomalous $WW\gamma$ coupling \cite{Ari:2016aac,Senol:2017uek}. As seen in Fig.~\ref{fig9}, $\left(P_{e^-},P_{e^+}\right)=\left(-80\%,+60\%\right)$ polarization has larger cross sections compared to other polarization configurations.

We have estimated using $\chi^2$ analysis with a systematic error to obtain the constraints on the anomalous coupling parameters at the 95$\%$ C.L.. $\chi^2$ function is defined by \cite{Hernandez:2019aaa,Koksal:2019mkn,Billur:2019mkn}:

\begin{eqnarray}
\label{eq.20} 
\chi^2=\left(\frac{\sigma_{SM}-\sigma_{NP}}{\sigma_{SM}\sqrt{\left(\delta_{st}\right)^2+\left(\delta_{sys}\right)^2}}\right)^2\,.
\end{eqnarray}

Here, $\sigma_{SM}$ is the cross section in the SM and $\sigma_{NP}$ is the cross section containing both the SM and new physics contributions. $\delta_{st}=\frac{1}{\sqrt {N_{SM}}}$ and $\delta_{sys}$ are the statistical error and the systematic error. The number of SM events is presented by $N_{SM}={\cal L}\times \sigma_{SM}$, where ${\cal L}$ is the integrated luminosity. 

The systematic uncertainty value has been taken into account in previous electron positron collider studies. The systematic uncertainty in the total cross section analysis for the process $e^-e^+\,\rightarrow\,t\overline{t}$ in electron positron collider is considered to be 3$\%$ and also the systematic uncertainty in determining the cross section has been reduced from 3$\%$ to 1$\%$ in the LEP \cite{Martinez:2003ert}. Since the ILC will be built in the coming years, it can be assumed that systematic uncertainties will be lower with the development of future detector technology. Taking into consideration the previous studies, we consider the systematic uncertainties of $0\%$, $1\%$ and $3\%$ in this paper. The measurements with small enough systematic uncertainties are needed to provide the required sensitivity for the new physics research. 

Also, in our study, we consider the effect of the uncertainty on the electron and positron beam polarization. There are very important to know the exact polarization because deviation from the expected polarization can fake contributions from the anomalous couplings.  The measurement of the polarization is performed by a polarimeter. The systematic uncertainties affecting the polarization measurements are (a) laser polarization, (b) detector linearity, (c) analyzing power calibration and (d) electric noise which led to up to an uncertainty of $\Delta P_e/P_e\backsim0.5\%$. For this reason, our analyses include polarization uncertainty of both the electron and positron beam polarization of $0.5\%$, $0.25\%$ or $0.1\%$. Also, the crucial point is that the systematic uncertainties can be significantly reduced when both beams are polarized \cite{Moortgat:2008yhn}. Here, we model the systematic uncertainty by adding uncertainties of the values of $\delta P_{e^+}$ and $\delta P_{e^-}$ to the systematic uncertainty introduced in $\chi^2$ analysis. 

The limits on the anomalous $c_{WWW}/\Lambda^2$, $c_{W}/\Lambda^2$ and $c_{B}/\Lambda^2$ coupling parameters at the 95$\%$ C.L. for the process $e^- e^+ \,\rightarrow\,\nu_e \overline{\nu}_e \gamma$ are given in Tables~\ref{tab5}-\ref{tab7}, depending on the polarization scenarios, the kinematic cuts and the systematic errors. The limits for each polarization scenario and systematic error become more sensitivity with increasing kinematic cuts (from Cut-1 to Cut-3d). The increment in systematic error causes a decrease in the sensitivity of all limits. The polarization $\left(P_{e^-},P_{e^+}\right)=\left(-80\%,+60\%\right)$ has the best limits for all anomalous coupling parameters. We can sort the polarizations corresponding to the less sensitive limit than the most sensitive limit: $\left(P_{e^-},P_{e^+}\right)=\left(-80\%,+60\%\right)$, $\left(-80\%,+30\%\right)$, $\left(P_{e^-},P_{e^+}\right)=\left(-80\%,0\%\right)$, unpolarized, $\left(P_{e^-},P_{e^+}\right)=\left(+80\%,0\%\right)$, $\left(P_{e^-},P_{e^+}\right)=\left(+80\%,-30\%\right)$ and $\left(-80\%,-60\%\right)$, respectively.

\begin{table}[H]
\renewcommand\thetable{V}
\caption{95\% C.L. constraints on the anomalous $c_{WWW}/\Lambda^2$ coupling for seven different polarization scenarios and three different cuts.}
\label{tab5}
\begin{ruledtabular}
\begin{tabular}{ccccc}
\multicolumn{2}{c}{} & \multicolumn{3}{c}{$c_{WWW}/\Lambda^2$ (TeV$^{-2}$)} \\
\hline
Polarization scenarios & Cuts & $\delta_{sys}=0\%$ & $\delta_{sys}=1\%$ & $\delta_{sys}=3\%$ \\ 
\hline \hline
\multirow{3}{*}{Unpolarized} 
 & Cut-1 & [-0.02469; 0.03157] & [-0.26568; 0.27260] & [-0.46264; 0.46962]\\ 
 & Cut-2 & [-0.02222; 0.02775] & [-0.16178; 0.17732] & [-0.28694; 0.28850]\\ 
 & Cut-3d & [-0.01317; 0.01137] & [-0.05263; 0.05083] & [-0.09042; 0.08863]\\ \hline
\multirow{3}{*}{$P_{e^-}=+80\%; P_{e^+}=0$} 
 & Cut-1 & [-0.04326; 0.05896] & [-0.27115; 0.28681] & [-0.47518; 0.49076]\\ 
 & Cut-2 & [-0.03209; 0.05010] & [-0.16315; 0.18015] & [-0.28871; 0.30669]\\ 
 & Cut-3d & [-0.02714; 0.02058] & [-0.06453; 0.05796] & [-0.10873; 0.10217]\\ \hline
\multirow{3}{*}{$P_{e^-}=-80\%; P_{e^+}=0$} 
 & Cut-1 & [-0.02213; 0.02902] & [-0.26081; 0.27069] & [-0.45707; 0.46691]\\ 
 & Cut-2 & [-0.01603; 0.02523] & [-0.15931; 0.17550] & [-0.28162; 0.28579]\\ 
 & Cut-3d & [-0.01201; 0.01103] & [-0.05104; 0.04995] & [-0.08831; 0.08821]\\ \hline
\multirow{3}{*}{$P_{e^-}=+80\%; P_{e^+}=-30\%$} 
 & Cut-1 & [-0.05386; 0.06596] & [-0.28380; 0.29592] & [-0.49570; 0.50785]\\ 
 & Cut-2 & [-0.03905; 0.05475] & [-0.16542; 0.18112] & [-0.29173; 0.30741]\\ 
 & Cut-3d & [-0.03256; 0.02559] & [-0.07287; 0.06375] & [-0.12185; 0.11274]\\ \hline
\multirow{3}{*}{$P_{e^-}=-80\%; P_{e^+}=+30\%$} 
 & Cut-1 & [-0.02084; 0.02876] & [-0.25714; 0.26814] & [-0.45296; 0.46415]\\ 
 & Cut-2 & [-0.01541; 0.02335] & [-0.15890; 0.17384] & [-0.28022; 0.28115]\\ 
 & Cut-3d & [-0.01059; 0.01026] & [-0.05075; 0.04902] & [-0.08694; 0.08651]\\ \hline
 \multirow{3}{*}{$P_{e^-}=+80\%; P_{e^+}=-60\%$} 
 & Cut-1 & [-0.06664; 0.08248] & [-0.30437; 0.32020] & [-0.53250; 0.54831]\\
 & Cut-2 & [-0.04774; 0.06701] & [-0.17029; 0.18954] & [-0.30103; 0.32025]\\ 
 & Cut-3d & [-0.04181; 0.03483] & [-0.08700; 0.08001] & [-0.14662; 0.13964]\\ \hline
\multirow{3}{*}{$P_{e^-}=-80\%; P_{e^+}=+60\%$} 
 & Cut-1 & [-0.01809; 0.02725] & [-0.25423; 0.26543] & [-0.44987; 0.46114]\\
 & Cut-2 & [-0.01368; 0.02294] & [-0.15461; 0.17186] & [-0.26149; 0.27773]\\
 & Cut-3d & [-0.00992; 0.00949] & [-0.05011; 0.04849] & [-0.08560; 0.08497]\\
\end{tabular}
\end{ruledtabular}
\end{table}

\begin{table}[H]
\renewcommand\thetable{VI}
\caption{Same as in Table~\ref{tab5}, but for the anomalous $c_{W}/\Lambda^2$ coupling.}
\label{tab6}
\begin{ruledtabular}
\begin{tabular}{ccccc}
\multicolumn{2}{c}{} & \multicolumn{3}{c}{$c_{W}/\Lambda^2$ (TeV$^{-2}$)} \\
\hline
Polarization scenarios & Cuts & $\delta_{sys}=0\%$ & $\delta_{sys}=1\%$ & $\delta_{sys}=3\%$ \\ 
\hline \hline
\multirow{3}{*}{Unpolarized} 
 & Cut-1 & [-0.62920; 0.01275] & [-1.19359; 0.57537] & [-1.77861; 1.15884]\\ 
 & Cut-2 & [-0.60630; 0.00808] & [-0.87582; 0.27652] & [-1.20368; 0.60422]\\
 & Cut-3d & [-0.51607; 0.00336] & [-0.56071; 0.04750] & [-0.63824; 0.12503]\\ \hline
\multirow{3}{*}{$P_{e^-}=+80\%; P_{e^+}=0$} 
 & Cut-1 & [-0.67621; 0.03672] & [-1.24673; 0.60774] & [-1.85909; 1.22114]\\
 & Cut-2 & [-0.64303; 0.02117] & [-0.94519; 0.28017] & [-1.28255; 0.61871]\\
 & Cut-3d & [-0.52486; 0.01025] & [-0.57685; 0.06225] & [-0.67305; 0.15846]\\ \hline
\multirow{3}{*}{$P_{e^-}=-80\%; P_{e^+}=0$} 
 & Cut-1 & [-0.62648; 0.01218] & [-1.19170; 0.56977] & [-1.77369; 1.15051]\\
 & Cut-2 & [-0.60586; 0.00745] & [-0.85119; 0.27414] & [-1.19143; 0.60312]\\
 & Cut-3d & [-0.51551; 0.00294] & [-0.55825; 0.04568] & [-0.63320; 0.12063]\\ \hline
\multirow{3}{*}{$P_{e^-}=+80\%; P_{e^+}=-30\%$} 
 & Cut-1 & [-0.70036; 0.04892] & [-1.28214; 0.63194] & [-1.91495; 1.26734]\\
 & Cut-2 & [-0.67458; 0.02775] & [-1.03499; 0.28990] & [-1.38746; 0.64607]\\
 & Cut-3d & [-0.53351; 0.01512] & [-0.59455; 0.07618] & [-0.70762; 0.18931]\\ \hline
\multirow{3}{*}{$P_{e^-}=-80\%; P_{e^+}=+30\%$} 
 & Cut-1 & [-0.62292; 0.01203] & [-1.19064; 0.56247] & [-1.77080; 1.14910]\\
 & Cut-2 & [-0.60164; 0.00686] & [-0.84463; 0.27125] & [-1.17396; 0.60244]\\
 & Cut-3d & [-0.51383; 0.00287] & [-0.55141; 0.04446] & [-0.63013; 0.12018]\\ \hline
 \multirow{3}{*}{$P_{e^-}=+80\%; P_{e^+}=-60\%$} 
 & Cut-1 & [-0.70856; 0.07846] & [-1.30543; 0.71244] & [-1.99565; 1.39655]\\
 & Cut-2 & [-0.68787; 0.04960] & [-1.04892; 0.33804] & [-1.39737; 0.71074]\\ 
 & Cut-3d & [-0.54867; 0.02584] & [-0.63056; 0.10775] & [-0.77922; 0.25648]\\ \hline
\multirow{3}{*}{$P_{e^-}=-80\%; P_{e^+}=+60\%$} 
 & Cut-1 & [-0.61953; 0.01184] & [-1.18989; 0.55458] & [-1.76596; 1.14843]\\
 & Cut-2 & [-0.59269; 0.00621] & [-0.81446; 0.26898] & [-1.16630; 0.60082]\\
 & Cut-3d & [-0.51126; 0.00279] & [-0.54646; 0.04400] & [-0.62063; 0.11920]\\
\end{tabular}
\end{ruledtabular}
\end{table}

\begin{table}[H]
\renewcommand\thetable{VII}
\caption{Same as in Table~\ref{tab5}, but for the anomalous $c_{B}/\Lambda^2$ coupling.}
\label{tab7}
\begin{ruledtabular}
\begin{tabular}{ccccc}
\multicolumn{2}{c}{} & \multicolumn{3}{c}{$c_{B}/\Lambda^2$ (TeV$^{-2}$)} \\
\hline
Polarization scenarios & Cuts & $\delta_{sys}=0\%$ & $\delta_{sys}=1\%$ & $\delta_{sys}=3\%$ \\ 
\hline \hline
\multirow{3}{*}{Unpolarized} 
 & Cut-1 & [-0.64133; 0.01152] & [-1.20754; 0.57143] & [-1.78503; 1.15431]\\
 & Cut-2 & [-0.62841; 0.00710] & [-0.91800; 0.28209] & [-1.25544; 0.61906]\\
 & Cut-3d & [-0.51991; 0.00322] & [-0.56391; 0.04683] & [-0.64058; 0.12350]\\ \hline
\multirow{3}{*}{$P_{e^-}=+80\%; P_{e^+}=0$} 
 & Cut-1 & [-0.65959; 0.03724] & [-1.23137; 0.60867] & [-1.84261; 1.21922]\\
 & Cut-2 & [-0.63845; 0.02175] & [-0.96433; 0.28327] & [-1.31001; 0.62531]\\
 & Cut-3d & [-0.52390; 0.01023] & [-0.57989; 0.06222] & [-0.67616; 0.15849]\\ \hline
\multirow{3}{*}{$P_{e^-}=-80\%; P_{e^+}=0$} 
 & Cut-1 & [-0.63852; 0.01106] & [-1.20529; 0.56921] & [-1.78381; 1.15150]\\
 & Cut-2 & [-0.58398; 0.00679] & [-0.87973; 0.27370] & [-1.22894; 0.61181]\\
 & Cut-3d & [-0.51165; 0.00297] & [-0.55462; 0.04594] & [-0.62986; 0.12116]\\ \hline
\multirow{3}{*}{$P_{e^-}=+80\%; P_{e^+}=-30\%$} 
 & Cut-1 & [-0.67507; 0.05084] & [-1.26546; 0.64122] & [-1.90191; 1.27767]\\
 & Cut-2 & [-0.65366; 0.02780] & [-0.96825; 0.28528] & [-1.31193; 0.63076]\\ 
 & Cut-3d & [-0.52781; 0.01518] & [-0.58887; 0.07624] & [-0.70182; 0.18918]\\ \hline
\multirow{3}{*}{$P_{e^-}=-80\%; P_{e^+}=+30\%$} 
 & Cut-1 & [-0.63781; 0.01054] & [-1.20383; 0.56219] & [-1.78175; 1.14741]\\
 & Cut-2 & [-0.56057; 0.00638] & [-0.79355; 0.26128] & [-1.19884; 0.59048]\\
 & Cut-3d & [-0.50595; 0.00286] & [-0.54441; 0.04533] & [-0.60901; 0.11995]\\ \hline
 \multirow{3}{*}{$P_{e^-}=+80\%; P_{e^+}=-60\%$} 
 & Cut-1 & [-0.68575; 0.07691] & [-1.31918; 0.70898] & [-2.00847; 1.39541]\\
 & Cut-2 & [-0.66447; 0.04750] & [-0.98933; 0.33127] & [-1.35370; 0.70325]\\ 
 & Cut-3d & [-0.54352; 0.02612] & [-0.62608; 0.10868] & [-0.77547; 0.25808]\\ \hline
\multirow{3}{*}{$P_{e^-}=-80\%; P_{e^+}=+60\%$} 
 & Cut-1 & [-0.61060; 0.01022] & [-1.17711; 0.56199] & [-1.76038; 1.13978]\\
 & Cut-2 & [-0.55819; 0.00637] & [-0.74075; 0.26019] & [-1.18652; 0.58950]\\
 & Cut-3d & [-0.50084; 0.00284] & [-0.54368; 0.04527] & [-0.60855; 0.11953]\\
\end{tabular}
\end{ruledtabular}
\end{table}

\section{Conclusions}

At the TeV scale, the LHC maybe provide new discoveries and valuable information.  On the other hand, it is usually agreed that the clean and precise environment of the linear colliders with respect to the LHC is ideally suited to the search for new physics and for determining precisely the underlying structure of the new interactions. Also, beam polarization is a key ingredient to the physics programme of future linear colliders. Polarized electron and positron beams in the linear colliders provide a valuable tool for stringent tests of the SM and for diagnosing new physics beyond the SM. In this respect, ILC is one step ahead of other linear colliders, because it offers different polarization options. Therefore, we examine the potential of the process $e^- e^+ \,\rightarrow\,\nu_e \overline{\nu}_e \gamma$ with unpolarized and polarized beams to study the non-standard $W^+W^-\gamma$ couplings at the ILC. 

The ILC will have polarized beams for both electron and positron beams to increase sensitivity to new physics and to improve precision measurement. Beam polarizations play a crucial role to increase the signal cross section while suppressing the unwanted background. A polarized beam provides a different viewpoint to test the SM and to research new physics beyond the SM. Observation of even the smallest signal which conflicts with the SM predictions would be a clue to prove the new physics. Proper selection of the electron and positron beam polarization may therefore be used to enhance the new physics signal. We have studied on the phenomenological aspects of the anomalous $W^+W^-\gamma$ couplings with the process $e^- e^+ \,\rightarrow\,\nu_e \overline{\nu}_e \gamma$ at the ILC. The total cross section and the limit analysis were performed according to the anomalous $c_{WWW}/\Lambda^2$, $c_{W}/\Lambda^2$ and $c_{B}/\Lambda^2$ coupling parameters that lead the deviations from SM. The cutflow has created by $p_T^\nu$, $|\eta^\gamma|$ and $p_T^\gamma$ cuts. According to this cutflow and polarization scenarios, the total cross sections were calculated against the anomalous coupling parameters. The polarization scenarios affecting the size of the total cross section in the largest cut (Cut-3d) have compared with each other. The ratios of total cross section to SM cross section on the anomalous $c_{WWW}/\Lambda^2$, $c_{W}/\Lambda^2$ and $c_{B}/\Lambda^2$ coupling parameters for polarization scenarios have determined and the contributions of the kinematic cuts to the signal have investigated. Using the $\chi^2$ analysis, the limits have obtained at 95$\%$ C.L. for the anomalous coupling parameters. If we look at the limits for each kinematic cuts corresponding to unpolarized and polarized beams in Tables~\ref{tab5}-\ref{tab7}, we can notice that the proper polarization of the leptons improve the limits on the anomalous couplings. We find that the polarization $\left(P_{e^-},P_{e^+}\right)=\left(-80\%,+60\%\right)$ provide the best limits on the all anomalous coupling parameters at the ILC.

Also, the sensitivity of the polarized positron is further improved compared to that of the unpolarized positron. There is an increase in the sensitivity on the $c_{WWW}/\Lambda^2$ coupling for Cut-3d of about $9.4\%$ between $\left(P_{e^-},P_{e^+}\right)=\left(-80\%,0\%\right)$ and $\left(P_{e^-},P_{e^+}\right)=\left(-80\%,+30\%\right)$ polarizations and about $15.7\%$ between $\left(P_{e^-},P_{e^+}\right)=\left(-80\%,0\%\right)$ and $\left(P_{e^-},P_{e^+}\right)=\left(-80\%,+60\%\right)$ polarizations, on the $c_{W}/\Lambda^2$ coupling for Cut-3d of about $1.4\%$ between $\left(P_{e^-},P_{e^+}\right)=\left(-80\%,0\%\right)$ and $\left(P_{e^-},P_{e^+}\right)=\left(-80\%,+30\%\right)$ polarizations and about $3.0\%$ between $\left(P_{e^-},P_{e^+}\right)=\left(-80\%,0\%\right)$ and $\left(P_{e^-},P_{e^+}\right)=\left(-80\%,+60\%\right)$ polarizations and on the $c_{B}/\Lambda^2$ coupling for Cut-3d of about $2.4\%$ between $\left(P_{e^-},P_{e^+}\right)=\left(-80\%,0\%\right)$ and $\left(P_{e^-},P_{e^+}\right)=\left(-80\%,+30\%\right)$ polarizations and about $3.3\%$ between $\left(P_{e^-},P_{e^+}\right)=\left(-80\%,0\%\right)$ and $\left(P_{e^-},P_{e^+}\right)=\left(-80\%,+60\%\right)$ polarizations.

The best limits have obtained for the anomalous coupling parameters by the CMS experiment at the CERN LHC \cite{Sirunyan:2019umc}. In order to prove the success of the study in this paper, our limits in Tables~\ref{tab5}-\ref{tab7} have compared with the experimental limits in Ref.~\cite{Sirunyan:2019umc}. It looks that the limits on the anomalous $c_{WWW}/\Lambda^2$, $c_{W}/\Lambda^2$ and $c_{B}/\Lambda^2$ coupling parameters obtained for all kinematic cuts and for all polarization scenarios are more sensitive than the limits in Ref.~\cite{Sirunyan:2019umc}. For Cut-1, Cut-2 and Cut-3d, the sensitivities of the limits corresponding to the polarization $\left(P_{e^-},P_{e^+}\right)=\left(-80\%,+60\%\right)$ on the anomalous $c_{WWW}/\Lambda^2$ coupling parameter are about 75, 90 and 165 times, that of the anomalous $c_{W}/\Lambda^2$ coupling parameter are about 115, 215 and 475 times and that of the anomalous $c_{B}/\Lambda^2$ coupling parameter are about 425, 680 and 1510 times better than the sensitivity of the limit in Ref.~\cite{Sirunyan:2019umc}. Even the limits corresponding to the polarization $\left(P_{e^-},P_{e^+}\right)=\left(+80\%,-60\%\right)$, which have the least sensitive limits among the polarization scenarios, have better sensitivity than the sensitivity of the limit in Ref.~\cite{Sirunyan:2019umc}; as for Cut-1, Cut-2 and Cut-3d about 20, 30 and 40 times on the anomalous $c_{WWW}/\Lambda^2$ coupling parameter, about 20, 30 and 55 times on the anomalous $c_{W}/\Lambda^2$ coupling parameter and about 60, 95 and 175 times on the anomalous $c_{B}/\Lambda^2$ coupling parameter, respectively. Our best limits for $c_{WWW}/\Lambda^2$, $c_{W}/\Lambda^2$ and $c_{B}/\Lambda^2$ coupling parameters were obtained for the polarization $\left(P_{e^-},P_{e^+}\right)=\left(-80\%,+60\%\right)$. The limits obtained for $c_{WWW}/\Lambda^2$, $c_{W}/\Lambda^2$ and $c_{B}/\Lambda^2$ coupling parameters for this polarization can be converted for $\Delta \kappa_{\gamma}$ and $\lambda_{\gamma}$ coupling parameters according to Eqs.~(\ref{eq.12})-(\ref{eq.13}). Therefore, the most sensitive limits on $\Delta \kappa_{\gamma}$ and $\lambda_{\gamma}$ couplings at the polarization $\left(P_{e^-},P_{e^+}\right)=\left(-80\%,+60\%\right)$ are converted; 

\begin{eqnarray}
\label{eq.21}
-3.14\times10^{-3}<\Delta \kappa_{\gamma}<5.63\times10^{-5}\,, \\
-7.45\times10^{-5}<\lambda_{\gamma}<11.23\times10^{-5}\,\, \text{for Cut-1}\,, \nonumber 
\end{eqnarray}
\begin{eqnarray}
\label{eq.22}
-2.94\times10^{-3}<\Delta \kappa_{\gamma}<3.21\times10^{-5}\,, \\
-4.45\times10^{-5}<\lambda_{\gamma}<7.47\times10^{-5}\,\, \text{for Cut-2}\,, \nonumber 
\end{eqnarray}
\begin{eqnarray}
\label{eq.23}
-2.58\times10^{-3}<\Delta \kappa_{\gamma}<1.44\times10^{-5}\,, \\
-4.09\times10^{-5}<\lambda_{\gamma}<3.91\times10^{-5}\,\, \text{for Cut-3d}\,. \nonumber 
\end{eqnarray}

These limits can be easily compare with the limits of the phenomenological studies in Table~\ref{tab1}. The $\lambda_{\gamma}$ coupling in Eq.~(\ref{eq.23}) appears to be more sensitive than the limits obtained for ILC, CLIC and CEPC in the phenomenological studies in Table~\ref{tab1}. But, we cannot say the same for the $\Delta \kappa_{\gamma}$ coupling in Eq.~(\ref{eq.23}). Because, although the positive limit of $\Delta \kappa_{\gamma}$ coupling in Eq.~(\ref{eq.23}) is more sensitive than the positive limits obtained for ILC, CLIC and CEPC in phenomenological studies in Table~\ref{tab1}, the sensitivity of the negative limit of $\Delta \kappa_{\gamma}$ coupling is lower than the others. In addition, considering the systematic uncertainties of $\delta_{sys}=1\%,3\%$, although the sensitivities of limits decrease, it is seen that they are better compared to the sensitivity of the limits in Ref.~\cite{Sirunyan:2019umc}.

As a result, we highlight that the sensitivities of the limits in this study are better than the sensitivity of the experimental limits reported for the LHC. Using polarized beams at the ILC to examine the anomalous $W^+W^-\gamma$ coupling through the process $e^- e^+ \,\rightarrow\,\nu_e \overline{\nu}_e \gamma$ provides great advantages for sensitivity studies by guaranteeing precise measurements.

\end{document}